\documentclass{article}
\usepackage{spconf,amsmath,graphicx}
\usepackage[dvipsnames]{xcolor}
\usepackage{tablefootnote}
\usepackage{booktabs}
\usepackage{multirow}
\usepackage{siunitx}
\usepackage{float}
\usepackage{hyperref}
\hypersetup{
    colorlinks=true,
    linkcolor=blue,
    filecolor=magenta,
    urlcolor=cyan,
}

\title{Depth Estimation from Monocular Images and Sparse radar using Deep Ordinal Regression Network}

\name{Chen-Chou Lo and Patrick Vandewalle}
\address{EAVISE, PSI, Dept.\ of Electrical Engineering (ESAT), KU Leuven \\
Jan de Nayerlaan 5, 2860 Sint-Katelijne-Waver, Belgium}

\begin{document}
%\ninept
%
\maketitle

\begin{abstract}
We integrate sparse radar data into a monocular depth estimation model and introduce a novel preprocessing method for reducing the sparseness and limited field of view provided by radar. We explore the intrinsic error of different radar modalities and show our proposed method results in more data points with reduced error. We further propose a novel method for estimating dense depth maps from monocular 2D images and sparse radar measurements using deep learning based on the deep ordinal regression network by Fu et al. Radar data are integrated by first converting the sparse 2D points to a height-extended 3D measurement and then including it into the network using a late fusion approach. Experiments are conducted on the nuScenes dataset. Our experiments demonstrate state-of-the-art performance in both day and night scenes.
\end{abstract}

\begin{keywords}
monocular depth estimation, radar, ordinal regression network, nuScenes
\end{keywords}

% INTRODUCTION
\section{Introduction}
\label{sec:intro}
A deep understanding of the outdoor 3D scene geometry is crucial to empower autonomous vehicles. This requires a very precise depth map of the vehicle environment, which can be either detected by depth sensors such as lidar or estimated through stereo or monocular RGB images. While stereo depth estimation algorithms predict a pixel-wise disparity map based on a calibrated stereo pair of images, monocular depth estimation algorithms estimate depth from only a single (monocular) RGB image. 
\begin{figure}[htb]
\begin{minipage}[b]{0.495\linewidth}
  \centering
  \centerline{\includegraphics[width=4.25cm]{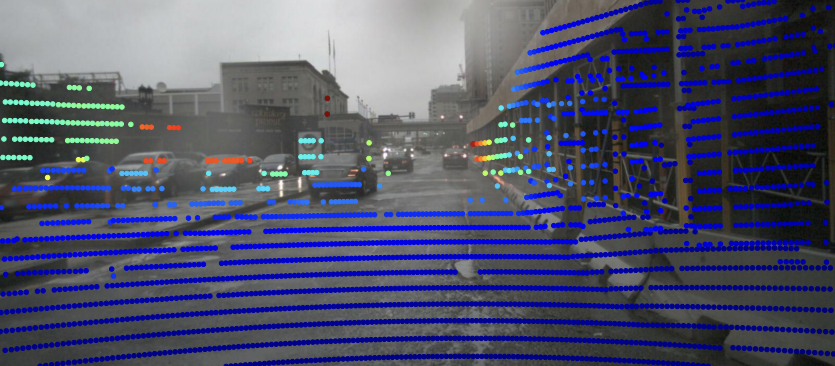}}
  \vspace{-0.1cm}
  \centerline{(a)}\medskip
\end{minipage}
\hfill
\begin{minipage}[b]{0.495\linewidth}
  \centering
  \centerline{\includegraphics[width=4.25cm]{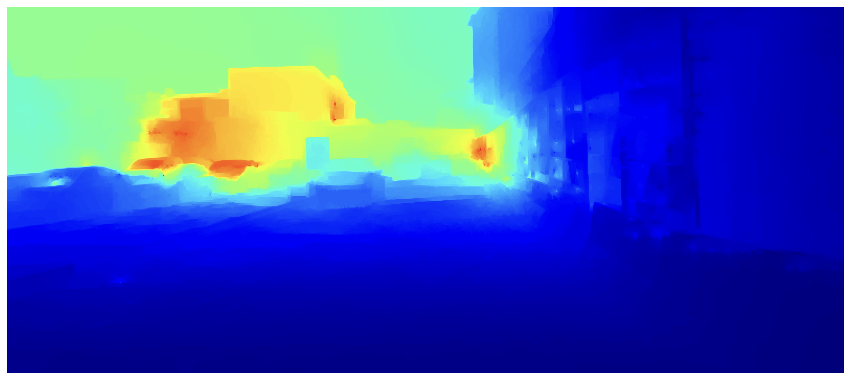}}
 \vspace{-0.1cm}
  \centerline{(b)}\medskip
\end{minipage}
\begin{minipage}[b]{0.495\linewidth}
  \centering
  \centerline{\includegraphics[width=4.25cm]{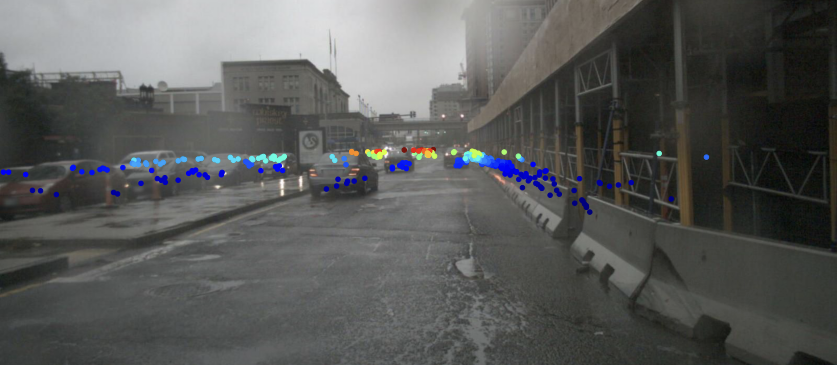}}
 \vspace{-0.1cm}
  \centerline{(c)}\medskip
\end{minipage}
\hfill
\begin{minipage}[b]{0.495\linewidth}
  \centering
  \centerline{\includegraphics[width=4.25cm]{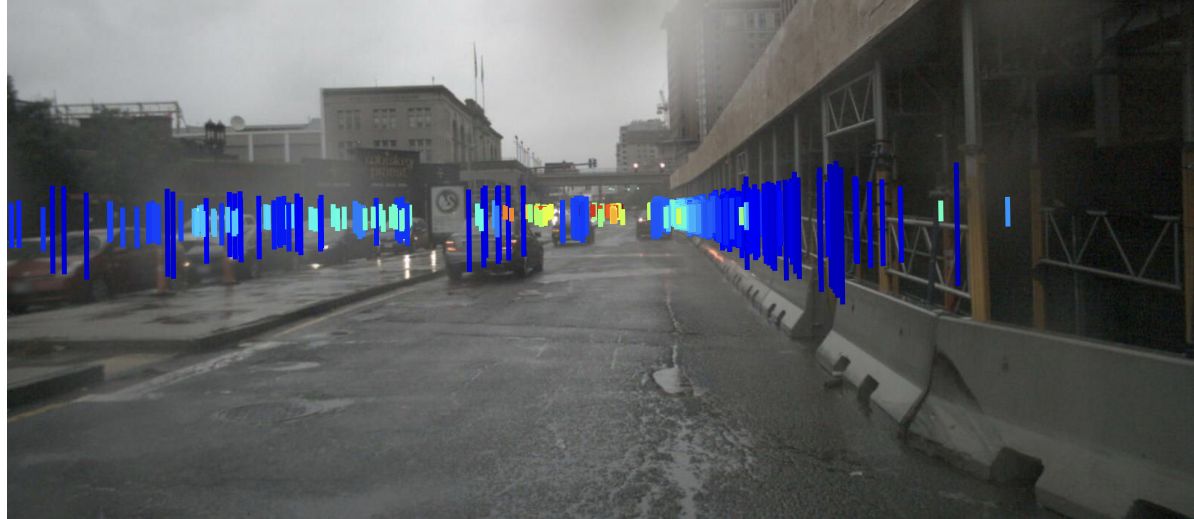}}
 \vspace{-0.1cm}
  \centerline{(d)}\medskip
\end{minipage}
\vspace{-4mm}
\caption{Samples from nuScenes dataset. (a) Projected sparse lidar; (b) interpolated dense depth; (c) 5-frames sparse radar; (d) 5-frames height-extended sparse radar. All the point sizes are dilated for better visualization.}
\vspace{-3mm}
\label{fig:sample_images}
\end{figure}

In recent years, the performance of depth estimation algorithms has significantly improved thanks to the development of deep neural networks for image-level feature retrieval and reconstruction. Many works have developed monocular depth estimation models \cite{eigen_1,eigen_2,DORN,bts,BANet,vip_deeplab}, leveraging different architectures and scales of features to improve the extracted representation and resulting in higher accuracy prediction. However, these monocular image-based approaches are limited by the lack of absolute depth information in a single RGB image and the ill-posed intrinsic character of the problem (an infinity of scenes could project to the same 2D image). Hence, many researchers \cite{s2d,self_s2d,SparseAD} have integrated corresponding lidar data along with RGB images to compensate for the missing absolute depth information and to improve the performance.

Although lidar can add a lot of relevant information about the 3D scene, it is also known to be sensitive to weather conditions and high-resolution sensors are very expensive. Meanwhile, radar has been used for commercial and military purposes for decades because of its robustness to different outdoor conditions and relatively low cost. However, due to its inherent sparseness, noise and limited field of view, radar is absent from most recently developed depth estimation algorithms and datasets. Recently, a new autonomous driving dataset was released including sparse radar data: nuScenes \cite{nuScenes}. This has allowed some recent works \cite{MDE_radar, MultiModalDE} to conduct experiments and propose a depth estimation model with radar data integrated.

\begin{figure*}
\vspace{-13mm}
  \includegraphics[width=\textwidth,height=4cm]{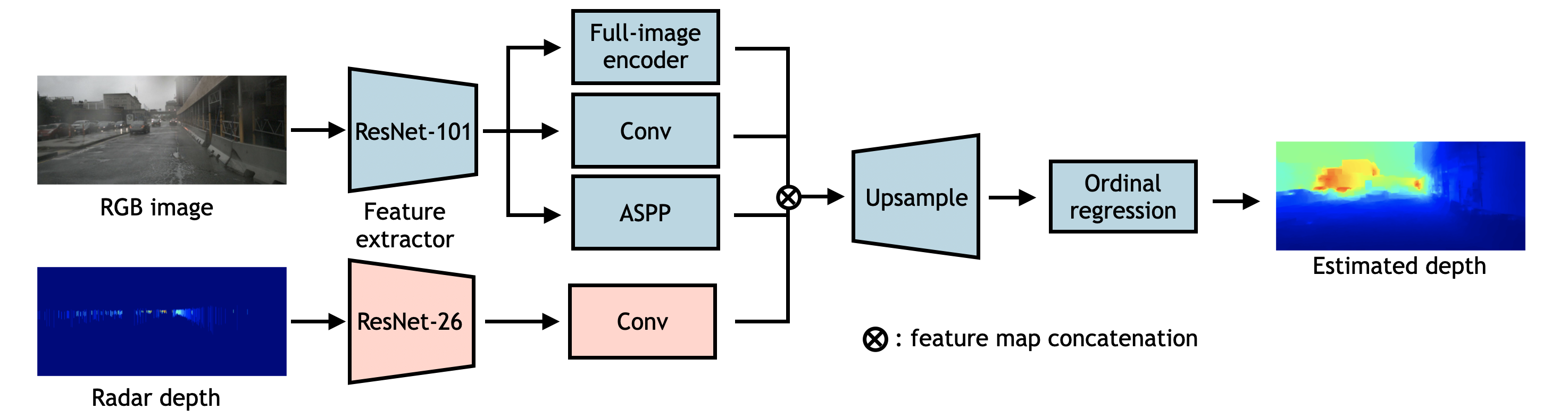}
  \caption{Illustration of the proposed architecture. Detailed description in Section \ref{subsec:architecture}.}
  \label{fig:arch}
\end{figure*}

In this work, we propose a method fusing height-extended multi-frame radar data with RGB images and experiment with both early fusion and late fusion integration into a state-of-the-art monocular depth estimation algorithm DORN \cite{DORN}. Through our experiments, we show (1) that enlarging the reach of sparse radar data by extending each point to a specific height improves the estimated depth accuracy, and (2) that depth estimation models, not designed for taking multi-modal input, can also benefit from sparse radar data with a proper fusion strategy and result in a superior prediction. Source code for our work is available online at
\href{https://github.com/lochenchou/DORN_radar}{https://github.com/lochenchou/DORN\_radar}.

% This paper is organized as follows: related works using RGB, lidar, and radar are reviewed in Section {\ref{sec:related_work}}. Our proposed method integrating radar with DORN is presented in Section {\ref{sec:method}}. Our experiments and results are discussed in Section {\ref{sec:experiments}}, and we conclude the paper in Section {\ref{sec:conclusion}}.

%
% RELATED WORK
\vspace{-2mm}
\section{Related work}
\label{sec:related_work}

\vspace{-2mm}
\subsection{Monocular depth estimation}
Recently, many improved depth estimation models have been proposed, motivated by the success of deep neural networks in image understanding. Eigen et al. \cite{eigen_1,eigen_2} proposed a multi-scale network to retrieve local and global feature maps stage-wisely to refine estimated depth. Fu et al. \cite{DORN} formulated depth learning as an ordinal regression problem (DORN) and showed faster convergence with ordinal regression loss. Lee et al. \cite{bts} proposed local planar modules to replace upsampling layers and reconstruct encoded feature maps back to full resolution depth. Aich et al. \cite{BANet} introduced bidirectional attention modules that derive from neural machine translation to utilize local and global feature maps in CNN layers to filter out ambiguity. Qiao et al. \cite{vip_deeplab} proposed a model to solve the inverse projection problem that jointly performs monocular depth estimation and video panoptic segmentation and restores a point cloud from image sequences.

\vspace{-2mm}
\subsection{Multi-Modal depth estimation}
In addition to estimating depth from RGB images, many works exploit the information from different modalities to improve the accuracy of output prediction. Because low-resolution lidar is cheap and easy to get, several researchers have proposed algorithms based on fusing RGB with sparse lidar. 
%Uhrig et al. \cite{sparseCNN} proposed sparse invariant CNNs that can consider the location of missing data during the convolution operation and extract a better representation from sparse input. 
Ma et al. \cite{s2d} integrated randomly sampled sparse lidar data as an additional channel to RGB images as input to a deep encoder-decoder network and further developed a self-supervised version \cite{self_s2d}. Compared with this early fusion strategy, Jaritz et al. \cite{SparseAD} proposed a late fusion method to handle sparse lidar and RGB images and improved the depth completion while accomplishing semantic segmentation. 

Radar has been used far more often than lidar as a sensing modality in practical military and aviation environments because of its robustness and cost. Lin et al. \cite{MDE_radar} conducted comprehensive experiments based on the depth completion network from \cite{s2d}, replacing lidar by radar data using different fusion strategies. Their results showed that the late fusion method, concatenating RGB and radar features after the encoder phase, has the best performance among all fusion methods. Siddiqui et al. \cite{MultiModalDE} proposed integration of radar and a monocular depth estimation model with an early fusion approach, and showed slightly enhanced results on a synthetic dataset but no improvements on the nuScenes dataset.

We believe that the poor performance reported in \cite{MultiModalDE} is caused by their early fusion method, directly inserting the noisy and sparse radar data as an extra channel in the input data. Therefore, we propose a different fusion approach, motivated by the late fusion method from \cite{SparseAD,MDE_radar}, as presented in Section {\ref{sec:method}}.

% Although a negative result is reported by \cite{MultiModalDE}, the reason would be that, compared with depth completion models which are designed for taking additional sparse depth as input, directly use the early fusion method with noisy and sparse radar data would not be enough for an MDE model to improve the prediction. Therefore, we address the issues of radar data and detail how we tackle it in Section {\ref{sec:method}}. The proposed late fusion method based on DORN is also presented in Section {\ref{sec:method}}.

\begin{figure*}[ht]
\vspace{-13mm}
  \includegraphics[width=\textwidth,height=4cm]{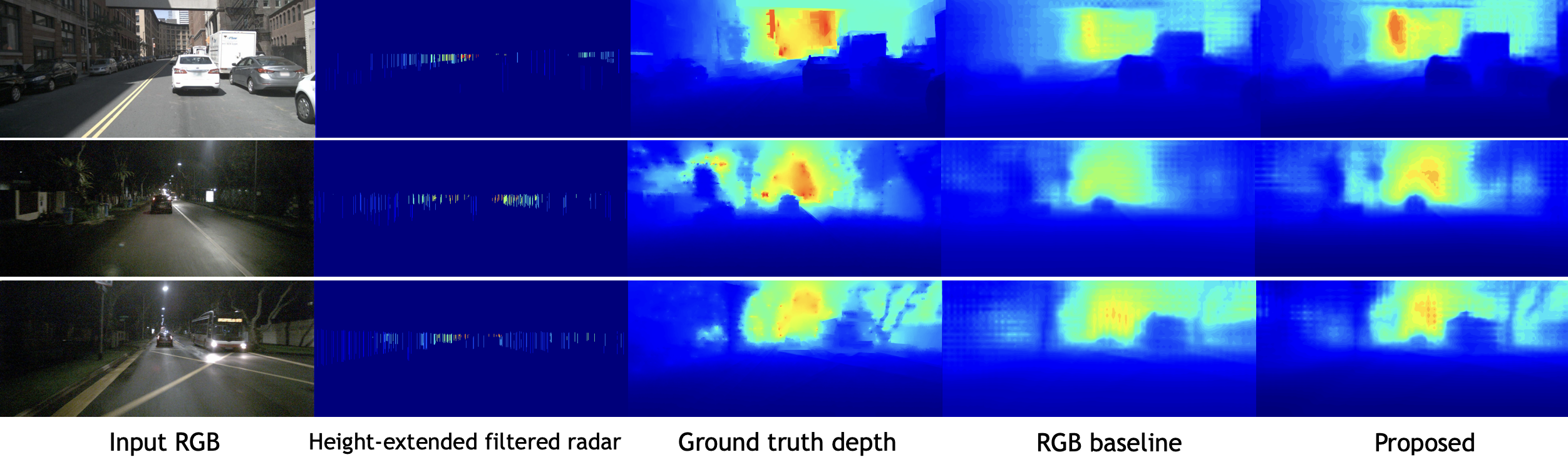}
  \caption{Examples of prediction on nuScenes in day (top) and night (middle and bottom) scenes. From left to right: input monocular RGB image; height-extended radar filtered via our proposed method; ground truth depth (interpolated via sparse lidar and a corresponding monocular RGB image); RGB-only baseline model (DORN); our proposed model.}
  \label{fig:result}
\end{figure*}

\vspace{-3mm}
\section{method}
\label{sec:method}
\vspace{-2mm}

In this section, we first introduce preprocessing methods for sparse radar data and present our proposed network architecture. Implementation details are described in the final subsection. 

\vspace{-2mm}
\subsection{Radar data preprocessing}
\vspace{-1mm}

Although radar is recorded in a similar point cloud format as lidar, three characteristics limit radar usage in depth estimation algorithms: 
\begin{enumerate}
  \vspace{-1.5mm}
  \item {\bf Sparseness}: the typical number of projected radar points in an image plane is in the order of ten points while lidar sensors deliver up to tens of thousands of points in an image plane.
  \vspace{-1.5mm}
  \item {\bf Noise}: as radar sensing is performed by detecting reflected radio waves, the resulting depth measurements are typically noisier than with lidar measurements.
  \vspace{-1.5mm}
  \item {\bf Field of view}: the acquired radar data in the nuScenes dataset is only two-dimensional, which means that we only have radar depth measurements in a single plane (parallel to the ground plane).
\end{enumerate}
For tackling the sparseness issue, we combine multiple frames of radar data into an image plane. For the limited vertical field of view, \cite{CRFNet} demonstrated that the accuracy of their proposed object detection model is improved with height-extended radar depth. Therefore, we extend the height of given multi-frame radar points from a single fixed height of 0.5m to a range of 0.25m to 2m height. 
Although such extended height radar depth can compensate for the limited vertical field of view, it also introduces potential erroneous measurements. Hence, we further follow the two-stage strategy in \cite{MDE_radar} to filter the noisy points, resulting in a height-extended multi-frame denoised radar depth.

\vspace{-3mm}
\subsection{Architecture}
\label{subsec:architecture}

\subsubsection{Deep ordinal regression network}
Deep ordinal regression network, DORN \cite{DORN}, is one of the state-of-the-art monocular depth estimation models. As illustrated in Fig.\ \ref{fig:arch}, the blue part in the proposed architecture (DORN) consists of a ResNet-101 \cite{ResNet} module as a dense feature extractor, followed by a scene understanding module including an atrous spatial pyramid pooling module \cite{ASPP} paralleled with a full image encoder and two $1\times 1$ convolutional layers to form the encoder part that converts an input image into a latent representation. For the decoder, the latent representation goes through a $1\times 1$ convolutional layer to adjust the channel size first, and then an upsampling layer is applied to decode back to the input image size. The main difference of the DORN network as compared with other models is the regression method. The use of the ordinal regression loss turns the depth estimation from a standard regression problem into a classification problem. Additionally, ordinal regression loss also takes ordinal correlation in depth values into account, which makes the model converge faster with higher accuracy.

\vspace{-3mm}
\subsubsection{DORN with radar}
Our goal is to integrate radar data into DORN, exploit the additional depth information and enhance the estimated depth especially under harsh weather conditions such as night scenes or rain scenes. We integrate the radar data with a late fusion manner as shown in Fig.\ \ref{fig:arch}. The dense feature extractor, ResNet-26, is randomly initialized and applied first to extract valuable features from the input sparse radar data. After adjusting the channel size by two $1\times 1$ convolutional layers, the radar feature map is concatenated with the RGB feature map and forms a scene understanding representation. The feature map further goes through an upsampling layer and regresses with the ordinal regression loss.

\newcommand{\ra}[1]{\renewcommand{\arraystretch}{#1}}
\begin{table}[H]
\vspace{-3mm}
\caption{Intrinsic error of raw and filtered radar data. We use interpolated dense depth for filtering (radar$^1$ for 5-frame radar; radar$^2$ for height-extended 5-frame radar; \#points for number of non-zero radar points in projected radar depth).}
\label{tab:radar_error}
\centering
\ra{1.2}
\begin{tabular}{cclll} 
\toprule
\multicolumn{1}{l}{modality} & threshold & $\delta$$_1$ $\uparrow$ & RMSE $\downarrow$ & \#points ($\%$) \\ 
\midrule
\multirow{2}{*}{radar$^1$}
& - & 0.433 & 22.829 & 205 (100) \\
% & \delta$_1$ & 1.0 & 4.489 & 86.8 (41.6) \\
& $\delta$$_2$ & 0.684 & 9.139 & 129.2 (61.8) \\
\cline{1-5}
\multirow{2}{*}{radar$^2$}
& - & 0.505 & 17.502 & 4510 (100) \\
% & \delta$_1$ & 1.0 & 3.889 & 2263.3 (49.7) \\
& $\delta$$_2$ & 0.715 & 7.468 & 3181.9 (69.9) \\
\bottomrule
\end{tabular}
\end{table}

\begin{table}[H]
\vspace{-3mm}
\caption{Comparison between different modalities and fusion methods (radar$^1$ for 5-frame radar; radar$^2$ for height-extended 5-frame radar).}
\label{tab:comparison_methods}
\centering
\ra{1.2}
\begin{tabular}{lcccc} 
\toprule
\multicolumn{1}{l}{modality} & fusion & $\delta$$_1$ $\uparrow$ & RMSE $\downarrow$ & AbsRel $\downarrow$ \\ 
\midrule
RGB & - & 0.872 & 5.382 & 0.117 \\
% \cline{1-2}
\multirow{2}{*}{RGB + radar$^1$ }
& early & 0.882 & 5.280 & 0.114 \\
& late & 0.884 & 5.222 & 0.107 \\
% \cline{1-2}
\multirow{2}{*}{RGB + radar$^2$ }
& early & 0.881 & 5.345 & 0.115 \\
& late & {\bf 0.887} & {\bf 5.194} & {\bf 0.107} \\
\bottomrule
\end{tabular}
\end{table}

\begin{table*}
\vspace{-13mm}
\centering
\caption{Comparison between different weather and time conditions. (radar$^2$ for height-extended 5-frame radar; radar$^3$ for $\delta$$_2$ filtered height-extended 5-frame radar)}
\label{tab:comparison_weathers}
\ra{1.2}
\begin{tabular}{@{}lcccccccccccccc@{}}
\toprule 
\multicolumn{1}{c}{\multirow{2}{*}{method}} & \multicolumn{4}{c}{$\delta$$_1$ $\uparrow$} & \multicolumn{1}{l}{} & \multicolumn{4}{c}{RMSE$\downarrow$} & \multicolumn{1}{l}{} & \multicolumn{4}{c}{AbsRel$\downarrow$} \\
\cmidrule{2-5} \cmidrule{7-10} \cmidrule{12-15}
& \multicolumn{1}{c}{overall} & \multicolumn{1}{c}{day} & \multicolumn{1}{c}{night} & \multicolumn{1}{c}{rain} & & \multicolumn{1}{c}{overall} & \multicolumn{1}{c}{day} & \multicolumn{1}{c}{night} & \multicolumn{1}{c}{rain} & & \multicolumn{1}{c}{overall} & \multicolumn{1}{c}{day} & \multicolumn{1}{c}{night} & \multicolumn{1}{c}{rain}\\ 
\midrule
% {Early fusion \cite{MDE_radar}} & 0.876 & 0.874 & 0.795 & - & & 5.628 & 5.574 & 6.723 & - & &  0.115 & 0.120 & 0.159 & - \\
%{Late fusion \cite{MDE_radar}} & 0.884 & 0.894 & 0.814 & - & &  5.409 & 5.271 & 6.402 & - & &  0.112 & 0.107 & 0.147  & - \\
% \midrule
{RGB} & 0.872 & 0.887 & 0.764 & 0.865 & & 5.382 & 5.150 & 7.122 & 5.637 & & 0.117 & 0.110 & 0.169 & 0.118\\
{RGB + radar$^2$} & 0.887 & 0.900 & 0.782 & 0.879 & & 5.194 & 4.978 & 6.861 & 5.488 & & 0.107 & 0.101 & {\bf 0.157} & 0.107\\
{RGB + radar$^3$} & {\bf 0.892} & {\bf 0.906} & {\bf 0.784} & {\bf 0.891} & & {\bf 5.082} & {\bf 4.845} & {\bf 6.856} & {\bf 5.268} & & {\bf 0.107} & {\bf 0.100} & 0.164 & {\bf 0.106}\\

\bottomrule
\end{tabular}
\end{table*}

\vspace{-2mm}
\subsection{Implementation details}
All the models we used in our experiments are implemented in a PyTorch \cite{pytorch} framework. To have a fair comparison, we used the same training settings for all experiments. The weights of ResNet-101 are initialized via the pretrained model on ILSVRC \cite{pretrained_weight}. We apply polynomial decay with a starting learning rate of 0.0001 and a power rate of 0.9 as the learning strategy. Batch size is set to 3, and momentum and weight decay are set to 0.9 and 0.0005 respectively. The discretization method for the ordinal regression layer is spacing-increasing discretization and we set the depth interval to 80, with a valid depth range from 1m to 80m. The number of training epochs is set to 25. To speed up training, we downsample RGB images, lidar depth, and radar depth from the original size of $900\times 1600$ to $450\times 800$. As the upper region has no depth values available, we further crop the RGB images, lidar, and radar depth into a shape of $350\times 800$, and use this as the training resolution. We accumulate the current radar frame with the previous 4 frames and extend each projected radar point to a height range of 0.25m to 2m. The ground truth dense depth for training is interpolated from sparse lidar and RGB images via the colorization method \cite{colorized}, and evaluation metrics are calculated with the ground truth sparse lidar. The input RGB images are normalized with the mean and standard deviation from Imagenet. While training, we further use data augmentation for RGB images to improve the robustness as follows: gamma contrast in range (0.9, 1.1); brightness adjustment in range (0.9, 1.1); color adjustment in range (0.9, 1.1); horizontal flipping with 0.5 probability. 
% Source code for our method is available online at \href{https://github.com/lochenchou/DORN_radar}{https://github.com/lochenchou/DORN\_radar}.

\vspace{-3mm}
\section{Experiments}
\label{sec:experiments}
We introduce the nuScenes dataset first and then show the evaluation result based on our proposed method. 
The evaluation metrics we use are following previous works \cite{MDE_radar, MultiModalDE}.

\vspace{-3mm}
\subsection{Dataset}
In our experiments, we make use of the recently released nuScenes dataset, a multi-modality autonomous driving dataset consisting of RGB images, sparse radar data, and 32-beam Velodyne lidar data. Scenes were captured in Boston and Singapore. nuScenes comprises 1000 driving scenes, each 20s long and fully annotated with 3D bounding boxes, in different locations and weather conditions. Among the 1000 scenes, 850 scenes are officially split as training set while the remaining 150 form the test set. We split the 850 training scenes in 765 training and 85 validation scenes, following the configuration in \cite{MDE_radar}. For the RGB images, we use the front camera only. In total, this results in 30731 training and 3418 validation pairs.

\vspace{-2mm}
\subsection{Intrinsic error of radar data}
To show the effectiveness of extending the height of radar data, Table \ref{tab:radar_error} demonstrates the intrinsic error calculated between projected radar depth and interpolated dense depth in training split. The average number of points of 5-frames radar is 215 with the intrinsic error of 0.350 for $\delta$$_1$ and 25.536 for RMSE. After extending the height, the average number of points increases with a factor of 22 and reaches 4510 points while the intrinsic $\delta$$_1$ and RMSE errors are better. Although extending the height could introduce noise, the overall intrinsic error is reduced. The same trend is visible after using $\delta$$_2$ as filter threshold, where the height-extended radar results in far more data points with lower error.

\vspace{-2mm}
\subsection{Overall comparison}
In Table \ref{tab:comparison_methods}, we compare the performance of both radar formats and fusion methods. It can be seen that all four models integrating radar as input yield better performance than the baseline RGB model (in all metrics). Among all the radar models, the height-extended radar with the late fusion model gives the best performance. The trend is correlated to the intrinsic result in Table \ref{tab:radar_error} that height-extended radar has more data points and less error. The late fusion method can also encode input radar into a better representation than early-fusion as also claimed in \cite{MDE_radar}.

\vspace{-2mm}
\subsection{Comparison in different weather conditions}
We further analyze the performance in day, night, and rain scenes with the baseline RGB model and the proposed late fusion with and without $\delta$$_2$ filtering on the height-extended radar in Table \ref{tab:comparison_weathers}. The result in the first two rows indicates that additional radar depths boost all metrics in all scenes, where the night scene has the highest performance gain since RGB images do not provide much information in such situations. With further filtering input radar by $\delta$$_2$ threshold via the prediction of the model itself and finetuning, the filtered radar can bring extra improvements compared with a raw noisy radar. Fig. \ref{fig:result} shows the output prediction from RGB-only baseline and our proposed model with RGB and filtered radar as input in day, night, and rain scenes. It can be seen that the prediction from our proposed method using self-filtered radar depth has more detail than the baseline in both day and night scenes.

\vspace{-5mm}
\section{Conclusion}
\label{sec:conclusion}
We have proposed a novel method for depth estimation from monocular RGB images and radar data. We combined height-extended multi-frame radar data with a monocular depth estimation model (DORN) comparing early and late fusion options. We have shown that a depth estimation model can benefit from a proper fusion method with preprocessed radar data. Our results demonstrated that the model could exploit radar information and enhance the prediction in day, night, and rain scenes.

\noindent\textbf{Acknowledgements}: This work was funded by a KU Leuven-Taiwan MOE Scholarship and Internal Funds KU Leuven.

% \cite{centernet,centerpoint}

\bibliographystyle{IEEEbib}
\bibliography{refs}

% \newcommand{\ra}[1]{\renewcommand{\arraystretch}{#1}}
% \begin{table}[H]
% \vspace{-3mm}
% % \caption{radar$^1$ for 5-frame radar; radar$^2$ for height-extended 5-frame radar; \#points for number of non-zero radar points in projected radar depth.}
% \label{tab:radar_error}
% \centering
% \ra{1.2}
% \begin{tabular}{cclll} 
% \toprule
% \multicolumn{1}{c}{radar} & \delta$_1$ \uparrow & RMSE \downarrow & \#points \\ 
% \midrule
% \multirow{1}{*}{raw} & 0.41 & 29.93 & 390.01  \\
% \multirow{1}{*}{height extended} & 0.46 & 21.52 & {\bf 9187.63} \\
% \multirow{1}{*}{signal expansion} & 0.43 & 28.04 & 6518.74 \\
% \multirow{1}{*}{proposed} & {\bf 0.51} & {\bf 17.27} & 8585.97 \\
% \bottomrule
% \end{tabular}
% \end{table}

\end{document}